\documentclass[twocolumn,prb]{revtex4}
\usepackage{graphicx,epsfig}

\def\be{\begin{equation}}
\def\ee{\end{equation}}
\def\jcr{J_{\rm cr}}
\def\S{{\cal S}}
\def\A{{\cal A}}
\def\P{{\cal P}}

\begin{document}

\title{The breakdown of the Nagaoka phase in the 2D $t-J$ model}

\author{E. Eisenberg$^{\dag\ddag}$, R. Berkovits$^{\dag\ddag\S}$,
David A. Huse$^{\dag}$ and B.L.
Altshuler$^{\dag\ddag}$}
\affiliation{$^{\dag}$Department of Physics, Princeton
University, Princeton, NJ 08544, USA\\$^{\ddag}$NEC Research
Institute, 4 Independence Way, Princeton, NJ 08540,
USA\\$^{\S}$Minerva Center and Department of Physics, Bar-Ilan
University, Ramat-Gan 52900, Israel}

\date{\today}

\begin{abstract}
In the limit of weak exchange, $J$, at low hole concentration, $\delta$,
the ground state of the 2D $t-J$ model is believed to be ferromagnetic.
We study the leading instability of this Nagaoka state, which emerges
with increasing $J$. Both exact diagonalization of small clusters, 
and a semiclassical analytical calculation of larger systems show 
that above a certain critical value of the exchange, 
$J_{\rm cr} \sim t\delta^2$, 
Nagaoka's state is unstable to phase separation. 
In a finite-size system a bubble of antiferromagnetic Mott insulator appears
in the ground state above this threshold. 
The size of this bubble depends on $\delta$ and scales as a power of
the system size, $N$. 
\end{abstract}

\maketitle

Recently, much interest was focused on the behavior of strongly correlated 
electron systems, which can not be explained by weak coupling 
perturbation theory. 
A variety of unusual phenomena such as, e.g. high temperature 
superconductivity and quantum magnetism, is believed to require a non 
perturbative description \cite{anderson}.
An important paradigm for the study of strongly interacting 
electrons in general, is the Hubbard model for interacting particles 
on a lattice. 
This is probably the simplest possible model that captures some of the behavior
of strongly correlated electrons. It is therefore widely used
to study various correlation driven effects, which are not described
by a perturbative approach. These include the metal-insulator (Mott)
transition, and the superconductivity of the high-$T_c$ compounds.

The Hubbard model was originally introduced in an attempt to describe 
quantum ferromagnetism of 
itinerant electrons in narrow band metals \cite{hubbard}. 
However, it is now well known that it is also a model for 
quantum antiferromagnetism.
The effective Hamiltonian which governs the low-energy 
behavior of the Hubbard model for a nearly half-filled
band in the $U\to\infty$ limit is the $t-J$ model
of itinerant fermions on the lattice
\be
H^{t-J}=-t\sum_{<ij>\sigma}a_{i\sigma}^\dagger a_{j\sigma}
+  c.c. +J\sum_{<ij>}\left ({\bf S}_i\cdot{\bf S}_j -\frac{1}{4}n_in_j \right )
.
\ee
The occupation number of each site
$n_i=a^\dagger_{i\uparrow}a_{i\uparrow}+
a^\dagger_{i\downarrow}a_{i\downarrow}$ can be either $0$ (a hole) or $1$
(a spin), since double-occupancy is forbidden by strong on-site Hubbard 
repulsion. 
The spin operators ${\bf S}_i$ are given in terms of the Pauli matrices 
$\sigma_{\alpha\beta}$ 
\be
{\bf S}_i=\frac{1}{2}\sum_{\alpha\beta}a_{i\alpha}^\dagger
{\bf \sigma}_{\alpha\beta}a_{i\beta}
~.
\ee
Since the spin exchange coupling constant $J\approx 4t^2/U$
is positive, the on-site Hubbard 
interaction translates to an anti-ferromagnetic (AFM) super-exchange, 
which favors an AFM correlated ground state. 

Nevertheless, in one extreme case this model has a ferromagnetic (FM)
ground state, known as the Nagaoka state \cite{nagaoka}. 
A fully polarized state
minimizes the total energy of a single hole in an 
otherwise half-filled band, at least in the limit $J\to 0$.
This is due to the hole kinetic energy which favors FM ordering.
The nature of the ground state for finite $J$ 
is therefore determined by the competition 
of this FM tendency with the AFM exchange. 

In this work, we study the ground state of the 2D $t-J$ model for 
low concentration of holes, as a function of
the AFM interaction strength $J$. We aim to identify
the leading instability of the Nagaoka state, i.e., the state which 
minimizes the energy for non-fully polarized states. Based on numerical 
evidence, we claim that the leading instability is towards
the creation of an AFM bubble, while the holes are confined to the FM 
region. The transition between Nagaoka state and this ``bubble'' state
is a first order transition \cite{emery}, including a jump
$\Delta S \gg 1$ in the total spin. We will concentrate on the
$t-J$ model, Eq. (1). However, our study is focused on the small $J$ 
regime, and thus our considerations apply equally to the Hubbard model.

We start by summarizing in some detail what is known and conjectured
about the nature of the 
$t-J$ model ground state. For half-filling,
where each site is occupied by one electron, no hopping is possible, and
the Hamiltonian reduces to the quantum Heisenberg Hamiltonian
with AFM spin exchange. 
Although the 2D Heisenberg model can not be solved analytically, there is 
strong numerical evidence, 
(obtained from exact diagonalization and Quantum Monte Carlo simulations)
that the ground state has long 
range AFM correlations at zero temperature\cite{rmp,rev}.
 
The presence of holes makes the picture more complicated.
Each hole hopping creates changes in the spin configuration, unless 
the spin polarization is uniform. The resulting
excitations in the system inhibit the hopping, since the hopping
probability is reduced by a factor proportional to the overlap between
the original and final spin wave functions. This effect results in 
narrowing the kinetic energy band, thus increasing
the kinetic contribution to the energy. The band width is maximized 
in the fully polarized state for which the spin configuration is unaffected 
by permutations of different spins. 
Thus, while the $J$ term in the Hamiltonian favors
an AFM ordering, the kinetic $t$ term favors a ferromagnetic (FM) state. 
The competition between these terms depends on the hole density and 
interaction strength $J/t$, and determines the physical properties of the 
$t-J$ model.

The question of charge carrier dynamics on an AFM background was 
extensively studied using the string picture, self-consistent Born 
approximation, and numerical studies \cite{rev}. 
For $J\gg t$ a single hole is unable to alter its
AFM surrounding, and its kinetic energy comes only through coupling to the
quantum fluctuations of the spin system. For moderately large $J$ ($J/t\approx
5-10$) the hole develops a spin-polaron around itself. Within this polaron,
the AFM order is suppressed by hole hopping. 
The size of the polaron increases with decreasing $J/t$.
The ``string'' picture emerges when this size is much larger than
the lattice constant. 
In this picture, the hole acts as a particle
with mass $1/t$ subject to an
effective potential resulting from the string of flipped spins created
by its hopping. This effective potential is a
linear potential with slope $J$. 
Accordingly, the size of the polaron scales like $(J/t)^{-1/3}$.
In this regime the hole movement
scrambles its surrounding spins, thus suppressing the local AFM order 
parameter, but does not yet create a local magnetic moment larger
than the spin-1/2 of a single hole.
\cite{br,ss}

As soon as $J/t$ decreases beyond some critical value ($J_c < 0.1$), the
nature of the ground state changes \cite{wa}. It turns out that lower energy 
is achieved by creating a FM polaron \cite{pok}, 
where not only the AFM ordering is destroyed, but also FM correlations
and a larger magnetic moment are formed in the vicinity of the hole. 
The radius of the polaron in this regime scales like $(J/t)^{-1/4}$.
In the limit $J\to 0$, the kinetic energy which favors FM ordering 
dominates the tendency towards AFM coupling of neighboring spins 
represented by the $J$ term, and the size of this polaron diverges. 
As a result, the ground state becomes fully polarized.
This statement was rigorously proven in the the celebrated paper of
Nagaoka \cite{nagaoka} for a single hole case.
It was shown that Nagaoka FM phase survives for sufficiently
small density of holes, even in the thermodynamic limit \cite{ber}.  

Thus, this apparently simple model exhibits a ferromagnet-antiferromagnet 
transition at zero temperature, as the interaction strength $J$ increases.
Very little is known about the properties of the system in the
transition region. Even the
dependences of the magnetization curve on $J$ and the
hole concentration $\delta$ are yet to be determined. 
In the following, we try to 
understand some features of the transition. We focus on the boundary regime 
of Nagaoka's phase, i.e., the transition from full to partial polarization.

Naturally, the question of the stability
of the Nagaoka fully polarized state with respect to switching on the AFM
exchange term, as well as finite hole density $\delta$, 
attracted much interest. 
Several variational wavefunctions were suggested in order to yield bounds 
for the Nagaoka stability region in $\delta$-$U$ plane (See 
\cite{hanisch} and references therein).
Most of these estimates were based on the belief
that the transition from Nagaoka state is 
continuous at $T=0$ \cite{hanisch}. In other words, 
it was implicitly assumed 
that the leading instability of Nagaoka's state is a single spin 
flip (SSF).
According to this picture, the transition to the AFM singlet ground state with 
increasing $J/t$ occurs 
gradually, through small incremental $\Delta S=1$ changes in the total spin.
However, it was suggested that for sufficiently low $\delta$ 
a phase separation instability, rather than a SSF one, might be 
relevant \cite{il,mhps}. 

Many authors have discussed the possibility of phase separation in the 
Hubbard and $t-J$ models \cite{il,viss,emery,21,ps1,ps2}.
For high concentration of holes (low electron density),
phase separation does occur for $J$ sufficiently large \cite{emery,21}.
However, the small $J/t$ case is still under debate. 
Some groups argue that the ground state phase separates for all values 
of $J/t$ for sufficiently low concentration of holes \cite{emery,ps1}. 
Others claim 
that there exists a critical value $J_{\rm ps}$ (estimates for its value 
vary between $J_{\rm ps}\sim 0.4t$ and $J_{\rm ps}\sim 1.4t$), such that for 
$J<J_{\rm ps}$ the ground state is uniform even for vanishing
hole density \cite{ps2}. 

In the following, we present evidence 
that the leading instability for low densities is indeed a phase separated 
state. We therefore claim that phase separation does occur for
small values of $J/t$, given that $\delta$ is sufficiently small. 
We further show that for a finite-size sample the transition in
the ground state between the Nagaoka phase and the 
phase separated state is discontinuous \cite{emery}, 
including a large change in the total spin $\Delta S >> 1$. 
Thus this transition can not be captured by 
the SSF variational studies. It is therefore reasonable to assume that 
better bounds for Nagaoka stability might require considering 
many simultaneous, rather than sequential, spin-flips.

We start by examining at the single hole case. It is straight forward to show 
that $\jcr$, the $J$ value needed to destabilize Nagaoka's state, 
for two spin flips {\it is smaller}
than for one spin flip \cite{il}. 
This can be demonstrated by deriving an effective Hamiltonian of
a single hole and flipped spins, in the background on the 
FM Nagaoka state. Since the number of flipped spins is presumably small, 
it is convenient to describe the
system in terms of the two types of ``particles'' - the hole and the 
flipped spins. The kinetic part of the Hamiltonian induces hopping of
the hole to its nearest neighbors, with an effective mass $1/(2t)$. The 
non diagonal part of the exchange interaction induces hopping of the
flipped spins to their nearest neighbors, with an effective mass $1/J$.
The ratio between the effective masses of
the spin and the hole is therefore $J/2t$, which is small around $\jcr$. 
Therefore, one can assume that the spin is static for calculating its effect 
on the hole energy, i.e., use Born-Oppenheimer-like adiabatic approximation. 
The diagonal part of the exchange term contributes the energy $-J/2$ per
flipped spin. In addition to these terms, one should take into
account the constraint that each site can be occupied by no more
than one ``particle'' (hole or flipped spin).
For that purpose, one can introduce effective,
infinitely strong, on-site repulsion between the particles.
There are other terms in the effective Hamiltonian resulting from
nearest neighbors interaction between the particles, but they do not
contribute to the energy in the leading order in $J/t$.
The Nagaoka state is destabilized as soon as
the $O(J)$ magnetic energy gained by the flipped spins overcomes this
kinetic energy gain. 

For a state with two spin flips, the kinetic 
energy increase depends on the distance between the spin flips. When they 
are far apart, the kinetic energy lost by the excluded area is about twice 
the value for a single flip. However, as the two spin flips get closer, 
the effective excluded area for the hole decreases.
When the distance is small compared to the hole wavelength, the 
excluded area is not changed much, and the kinetic energy increase is about 
the same as for one flip. Thus, there is an effective attraction between 
the flipped spins, and they bind together, with total 
energy similar to that of a single flip. On the other hand, 
the magnetic energy is simply proportional to the number of
spin flips, and does not depend on the distance 
between the spin flips, as long as they are not nearest neighbors.
Thus the magnetic energy for two flipped spins, bound or not,
is twice as much as the energy for a single flip. 
Therefore, $\jcr$, for which the magnetic energy balances the kinetic 
energy increase, is lower for a state with two flipped spins.
One concludes that, for a single hole, the transition between
the maximum spin
Nagaoka state and a lower spin state involves a spin jump $\Delta S>1$.

In order to describe the leading instability of the Nagaoka 
ferromagnet with  respect to switching on the $J$ term
one should, therefore, determine the number of spin flips which minimizes
$\jcr$.
We start with presenting results of
exact diagonalization for small rectangular $a\times b$
clusters. We choose periodic boundary conditions when the large axis of the
cluster has even number of sites ($a=2n$), and anti-periodic boundary 
conditions otherwise ($a=2n+1$). 
Under these conditions the $J=0$ ground state is always fully polarized. 
For each of the following clusters we diagonalize exactly
the full many-particle Hamiltonian, for different spin sectors and
various $J$ values. We determine $\jcr$ and the value of the ground 
state spin for $J=\jcr +$.
It was already pointed out \cite{ak},
based on exact diagonalization results,
that for big enough clusters, large number
of spins may flip together. We present here a more systematic study of
finite clusters, showing a large spin jump $\Delta S>1$.
For the one hole case, the biggest cluster studied was
a $6\times 4$ torus. The effective size of the full Hilbert space can
be substantially reduced by excluding the doubly occupied states and using
spin and translational symmetry. For $6\times 4$ torus, this reduces the 
problem to the diagonalization of a $1352078\times 1352078$ matrix.
We employed the Lanczos algorithm to carry out this diagonalization.
The results for the different clusters are summarized in 
table I. Extension of these exact diagonalization studies to larger 
systems or more than one hole 
is limited by the size of the Hilbert space.

\begin{table}[t] 
{ 
\begin{tabular}{|c|c|c|c|c|c|}
Cluster      & $\Delta S$ & $\jcr$ & Cluster & $\Delta S$ & $\jcr$ \\ \hline
$2\times 8 $ &  $3$      & $0.0225$& $4\times 4$&  $4$    & $0.0629$ \\
$2\times 10$ &  $4$      & $0.0134$& $4\times 5$&  $6$    & $0.0398$ \\
$2\times 12$ &  $5$      & $0.0086$& $4\times 6$&  $6$    & $0.0269$
\end{tabular}
\caption{\small Exact diagonalization results for the breakdown on Nagaoka's
state for the case of one hole. $\Delta S$ is the jump in the 
ground state spin at the transition.}}
\label{tab1}
\end{table}

We did repeat the calculation 
for two holes on the above clusters. It is well known that for two holes 
on a torus the ground state is a singlet even for $J=0$ \cite{doucot,ak}. 
This is due
to the fact that a slowly varying, locally polarized, spin background
creates the effect of a fictitious  flux, which minimizes the kinetic
energy of the two holes \cite{eb}. As it was done for one hole, 
we chose the boundary conditions 
to overcome this effect. However, in all cases studied,
a ground state with small nonzero spin was never observed.
We always dealt with an abrupt transition to the singlet $S=0$ state, 
$\Delta S$ being as large as the largest possible spin, $S_{\rm max}$, 
even larger than the jump in the spin for the one hole case.   
This, however, is a particular feature of systems with two 
holes, which might not be relevant for the behavior of large systems. 

These exact results show that the number of spins which flip at 
the transition is significantly larger than one, and it increases with 
the system size. They indicate that the breakdown of Nagaoka phase might 
indeed occur through an abrupt spin change.
It looks interesting to understand how $\Delta S$ scales with
system size. Besides, it is not clear to what extent are the results
specific for a particular small number of holes ($1$ or $2$), i.e.,
extension
of the results to more holes is required. Clearly, exact diagonalization 
is not an adequate tool to clarify these questions. 

To get an insight into the behavior of larger systems, 
we use the spin-hole coherent state 
path integral formalism. We follow the derivation presented in 
\cite{auerbach}. A semiclassical approximation for the
ground state is obtained through the formal large $S$ expansion. 
In the classical 
limit, $S\to\infty$, the spins are frozen (i.e., the spin part of the 
coherent state path integral is time-independent).
The energy is determined solely by the interplay 
between hole dynamics and the classical AFM interaction between the spins.
In essence, this limit deals with a classical spin field, interacting with 
the quantum holes. Clearly this approximation can not capture the
full complexity of the exact wave function, as it ignores dynamical
corrections to the spin background as introduced by, e.g., the dressed
holes. Thus, some of the exotic phases suggested for the $t-J$ model are
essentially not taken into account in this approximation. However,
in some cases the instability towards these phases can be described
within the semi classical approximation. For example, the
incommensurate AF, or spiral phase \cite{doucot,ak},
with dressed holes, supercedes the Nagaoka
phase for two holes. Within the semi classical approximation, ignoring
the dressing of the holes, the
incomensurate AF background becomes degenerate with the Nagaoka phase.
Thus one can expect to find signs for the emergence of complex spin
structures even within this simplified approach.
The same approach was previously used to study the formation of the 
FM polaron. Here we use the same technique 
to study the extreme low $J$ case $J=\jcr +$, at the breakdown of Nagaoka
phase, where the FM region occupies most of the system. 
We use a Monte-Carlo algorithm to find the spin configuration 
which minimizes the sum of the energy of the holes and the 
magnetic energy of the (classical) spin field.
 
\begin{figure}
\includegraphics[width=6cm]{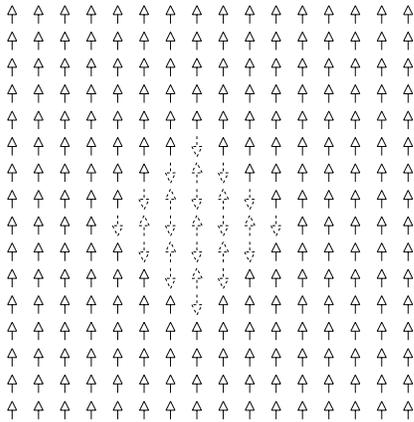}
\caption[]{ 
The spin configuration of the leading instability as obtained from
a semi-classical calculation. The hole density is $25/256$. 16 spins
flip in a slanted square configuration. The hole density vanishes 
inside the AFM bubble, while being approximately constant in the FM
region.
}
\label{f1}\end{figure}

We study different lattice sizes (up to $16\times 16$)
and various number of electrons close to half-filling ($0<\delta<0.12$). 
In all cases, we find the same behavior:
(a) all the spins align co-linearly
(b) we do not see any signature suggesting the emergence of 
exotic spin configurations (spiral, canted, etc.) 
for these small values of $J$, and most importantly (c) 
the uniform Nagaoka's state breaks down into a 
phase separated state, with a hole-rich FM
region and an AFM bubble with no holes inside. The
size of the bubble at the transition is large relative to the lattice 
constant, giving a big jump in the spin.
As an example we present in figure \ref{f1} results for a 
$16\times 16$ lattice with $25$  holes. 
These results, for different lattice sizes and densities, 
suggest that Nagaoka state breaks down by forming 
an AFM bubble, whose size is large compared to the lattice constant.
Contradicting evidence were recently presented  
regarding the existence of phase separation for small $J$ \cite{ps1,ps2}.
We find that the above described phase separated state
seems to be stable even for $J<<t$, 
provided that the hole density $\delta$ is sufficiently small,
in agreement with Ref. \cite{emery,ps1}. The stability of the phase
separated state for larger $J$, and its relation to the above discussed
phase, is beyond the scope of this study.

Motivated by these results, we turn to study the stability of 
the Nagaoka phase with respect to phase separation. That is, we compare the 
energy of the Nagaoka state with that of a system with an AFM ordered
domain. To first approximation, the AFM domain acts as an infinite potential
barrier, i.e. the holes are confined to the FM region. 
Due to this confinement, 
the kinetic energy of the holes increases with the size of the AFM domain.
On the other hand, the AFM domain contributes a magnetic term to the
energy, which also increases with its size. We estimate the magnetic energy 
per site in the AFM domain by the value obtained in an infinite Heisenberg
system, 
\be
E_{\rm mag}=- \alpha J (\S+\P/2),
\ee
where $\S$ is the area (number of sites) of the
AFM domain, $\P$ is its perimeter, and $\alpha$ is the energy per site
of the ground state of the 2D Heisenberg model 
(Spin wave theory gives $\alpha=1.1705$, 
which is in a good agreement with the numerical result
\cite{rmp})
$\alpha=1.169$. 
An estimate to the increase in the holes' kinetic energy for a large system
is obtained through the well known Weyl formula for the density of states. 
The number of levels up to an energy $E$ 
for a free particle in a continuous 2D domain with area $\A$ and
boundary perimeter $\P$ is given by 
\be
\langle N(E)\rangle
\sim \frac{1}{4\pi}(\A E - \P\sqrt{E} +{\cal K})
~,
\ee
where Dirichlet boundary conditions are assumed, and $\cal K$ is a constant
term containing information on the geometry and topology
of the domain \cite{book}. The units are chosen such that $\hbar=2m=1$.
Equivalently, for the Hubbard (or $t-J$) model, the same formula
holds near the bottom of the band, where the energy is in units of $t$
and the bottom of the band is taken as the energy zero.
In the following $t$ is taken as the unit of energy, and 
the lattice constant is taken as the unit of length.
The average $\langle \rho (E)\rangle$ of
the density of states (DOS) $\rho (E)\equiv\partial N(E)/\partial E$,
is thus given by
\be
\langle\rho(E)\rangle\sim \frac{\A}{4\pi}-\frac{\P}{8\pi\sqrt{E}}.
\ee

In the Nagaoka FM state, the available domain for the holes is the whole 
torus $\A=N$. The DOS is therefore just the familiar $\A/4\pi$ term and the 
energy is given by 
\be
\label{enag}
E_{\rm Nag}=2\pi N \delta^2 = 2\pi N_h^2/N
~,
\ee
where $N_h$ is the number of holes and  
$N$ is the total number of sites. As soon as 
the AFM bubble is formed, the energy increases due to two reasons. 
First of all, the available area is reduced to 
$\A=N-\S$, and therefore the denominator
in Eq. (\ref{enag}) decreases. Due to this fact, 
the total energy increases by the factor
$N/(N-\S)$. Another contribution comes from the boundary term which reduces 
the DOS even further. Bearing in mind to compare the 
energy increase with the magnetic energy which is $O(\S)$, we realize that
the increase due to the boundary term $O(\P)$ divided
by $\S$ becomes singular ($\sim \S ^{-1/2}$)
for small $\S$. We will see below that the transition occurs for
$\S$ values much smaller than the system size. 
As a result, boundary terms are important even for large $N$.
The topological $\cal K$ term also changes due to the AFM barrier. However,
it can be checked explicitly that it does not affect the $N\to\infty$
asymptotic behavior.

\begin{figure}
\includegraphics[width=6cm]{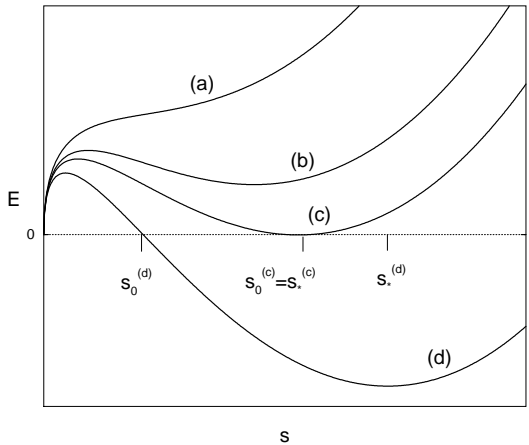}
\caption[]{ 
An illustration of the total energy (\ref{energy}) for different values 
of $j$. (a) No meta-stable state. (b) A meta-stable state appears
.
(c) The transition point. (d) $J>J_{\rm cr}$.
 }
\label{illus}\end{figure}

Using the above averaged DOS, one can calculate the total energy of holes 
as a function of their density. Adding the magnetic energy,
the total energy relative to the Nagaoka energy is given by
\begin{eqnarray}
\label{energy}
E(\delta,s)& =& 2\pi\delta^2 N \left (\frac{s}{1-s}
	+\frac{2p}{3\sqrt{\pi\delta N}(1-s)^{3/2}}\right )\nonumber \\
& &	-\alpha JN\left (s+\frac{p}{2\sqrt{N}}\right )
~,
\end{eqnarray}
where we introduced the normalized quantities 
\be
p\equiv\P/\sqrt{N}, \quad s\equiv\S/N.
\ee 
It follows from the dimensional analysis that $p=C\sqrt{s}$, where
$C$ is a dimensionless constant depending on the shape of the bubble. 
Although we deal with a lattice 
problem, in the limit we are considering of low doping and
large bubbles, the perimeter term from the hole kinetic energy
dominates.  Since the dispersion relation for low-energy holes
is asymptotically isotropic, this produces an isotropic perimeter
energy so the lowest energy bubbles are circular ($C=2\sqrt{\pi}$).
It is straight forward to see that given the grand canonical potential,
$G=E(\delta,s)-N\mu\delta$, with a fixed chemical potential $\mu$, that
the system goes through a first order phase transition. 
This transition occurs at $\jcr=\mu^2/(8\pi\alpha)$,
where the density jumps from $\delta_-=\mu/(4\pi)$ at $J=\jcr -$ to
$\delta_+=0$ at $J=\jcr +$. In the same time $s$ jumps from $s=0$ 
(Nagaoka phase) below the transition, to $s=1$ (antiferromagnet)
above the transition.  If the density is instead held
between these two values, the ground state, ignoring
surface terms, is a phase separated state, which is a mixture of the
two phases. In the following we study the formation of the 
AFM bubble in a finite system, given the surface terms.
The energy (\ref{energy}) as a function of $s$
is presented in figure \ref{illus} for various values of $J$.
At the origin,
$E(0)=0$, and the energy increases with $s$. For sufficiently large $J$, 
the function has a maximum at which the energy is positive
and a (local) minimum at $s=s_*$ (see Fig.~\ref{illus} curve (b)). 
When $J$ increases beyond a critical value $J>\jcr$, the value 
of the energy at this minimum becomes negative,
the local minimum at $s_*$ becomes the global minimum of the function.  
The function crosses the $x$-axis at $s=s_0<s_*$ 
(see Fig.~\ref{illus} curve (d)).
For $s>s_*$ the function again increases. 
Therefore, as long as $J<\jcr$, the global energy minimum is obtained at
$s=0$, corresponding to Nagaoka FM state, although a meta-stable 
state with an AFM bubble does exist for a range of $J$ near, but below $\jcr$.
As $J$ increases beyond $\jcr$, the global minimum shifts to $s_*$.
It is convenient to introduce the variable $j$ 
\be
j=\frac{J-J_0^{\rm cr}}{J_0^{\rm cr}};\quad J=2\pi\delta^2/\alpha (1+j)
\equiv J_0^{\rm cr}(1+j) 
\ee
$J_0^{\rm cr}$ is the value of $J_{\rm cr}$ for an infinite size system.
Thus, $j$ measures the relative distance of $J$ from the 
value of $\jcr$ obtained ignoring boundary terms.
In terms of this variable, one can express the large $N$ asymptotics of 
$s_0$ (the value of $s$ where the energy vanishes) and $s_*$, which minimizes
the energy.
\be
s_0=C^2(A/j-1)^2/4N;\quad A=4/(3\sqrt{\pi\delta})-1
\ee
\be
s_*=1-\frac{1}{\sqrt{1+j}}
\label{emag}
\ee
note that $A$ depends only on the hole density $\delta$.
Thus, for $j > 0$, the minimal size of the droplet needed to 
destabilize 
Nagaoka's state remains finite in the $N\to\infty$ limit ($S_0\equiv
s_0N=O(1)$), 
while the global minimum is obtained when the droplet is of macroscopic 
size ($S_*\equiv s_*N=O(N)$).
Comparing these two expressions, one immediately sees that the  
transition occurs for $j\sim O(N^{-1/3})$. 

\begin{figure}
\includegraphics[width=8cm]{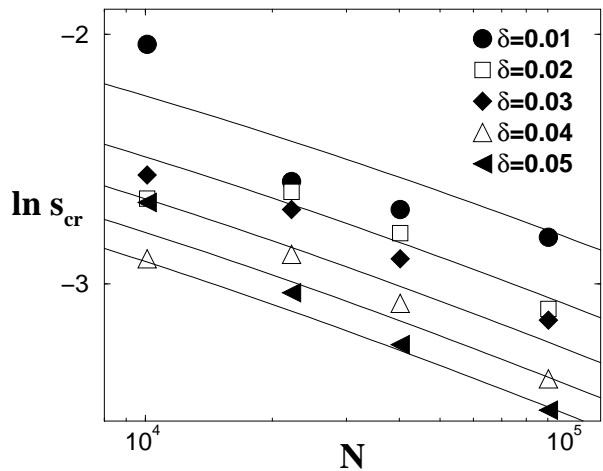}
\bigskip
\caption[]{
The area of the AFM bubble at the transition point, $s_{\rm cr}$, 
normalized by system size, 
as a function of $N$ for various $N$ values.
The solid lines are the asymptotic expansion up to order $N^{-1}$
for the same five densities (top to bottom)
.
}
\label{f3}\end{figure}

In this regime $s_0$ and $s_*$ are connected with $j$ via the equations
\be
j=2s_*+\frac{CA}{4\sqrt{Ns_*}}
\ee
\be
j=s_0+\frac{CA}{2\sqrt{Ns_0}}
\label{szero}
\ee
The last equation has a solution provided that $j>j_{cr}=3(C^2A^2/16N)^{1/3}$.
The transition point $J_{\rm cr}$ is therefore
\be
\label{jcr}
J_{\rm cr}=\frac{2\pi\delta^2}{\alpha}\left ( 1+3((CA)^2/16N)^{1/3} \right )
\ee
Beyond the transition point $J>J_{cr}$, and for some $s$ the 
energy of the bubble is negative and therefore the FM state is unstable.
At the transition point, the area of the optimal AFM bubble is
\be
\label{s0}
\S_0=Ns_0=Ns_*=(CAN/4)^{2/3}
\ee 
For $j<j_{cr}$ Eq.~(\ref{szero}) has no solution
and the energy is minimized by Nagaoka state. However,  
the energy function does have a local minimum at $s_*$ for $j>j_{\rm ms}
=2^{-1/3}j_{cr}$,
and thus a large meta-stable bubble can be created. The size of the 
meta-stable bubble at $j=j_{\rm ms}+$ is 
\be
S_*=s_*N=(CAN/16)^{2/3}
\ee

In fact, it can be shown that the $N^{2/3}$ power law follows from
a very general argument. Whenever phase separation occurs between two phases, 
with densities $\delta_1$
and $\delta_2$, the fraction of the first phase in the phase separated 
ground state of an infinite system at a fixed density 
$\delta_1<\delta<\delta_2$
is given by $(\delta_2-\delta)/(\delta_2-\delta_1)$. Thus,
for densities very close to $\delta_2$ the size of the bubble of the
first phase is arbitrarily small. However, the surface 
energy term raises the energy of the phase separated state 
as compared to the uniform state, and thus finite size corrections arise.
The energy cost to create a boundary between the two phases, is
proportional to (in $d$ dimensions) $V^{(d-1)/d}$, where $V$ is the 
size of the bubble. This energy makes it favorable to retain the uniform
phase even at densities slightly smaller than $\delta_2$. The formation
of a bubble becomes energetically favorable when the density is shifted
away from the (infinite system) transition point, such that the difference
in the volume energy of the uniform state as compared to the phase-separated 
state overcomes the surface term. In the generic case, this difference
is proportional to $V^2/N$ where $N$ is the system size, and $V$ is the
size of the optimal bubble. Comparing the two
energies, one observes that at the transition $V\sim N^{d/(d+1)}$.

We performed a numerical study to validate the above calculations. The 
single-particle spectrum of a tight binding square lattice model with 
periodic boundary conditions and an excluded domain, was
calculated  for different sizes of the excluded domain.
We then calculated the ground state kinetic energy 
(which is the sum of the lowest $m$ eigenvalues, 
where $m$ is the number of holes), and compared the increase
of kinetic energy as a result of the excluded bubble with the
gain in magnetic energy. For each size of excluded domain, 
a minimal value of $J$ is needed to balance the increase of kinetic
energy, thus stabilizing the AFM bubble. $\jcr$ is determined as
the minimum of these $J$ values, i.e., the lowest value of $J$ 
which allows for a stable bubble. The size of the bubble
at this minimum is $S_0$. The area of the stable bubble at 
$J=\jcr +$ is presented in figure \ref{f3} as a function of
$\delta$ and $N$.  It turns out that the leading asymptotic 
dominates only for a very big system, beyond numerical capabilities.
We therefore calculated the sub-leading corrections $O(N^{-2/3})$ and
$O(N^{-1})$ and compared the numerical results with this asymptotic
expansion. The agreement  with the
analytical result is quite good for sufficiently large lattices, 
and it seems that the results converge
to the asymptotic estimate as the system size increases, indicating that
corrections due to deviations from the free particle band shape are
not important for $\delta\leq 0.05$.

\begin{figure}
\includegraphics[width=7.3cm]{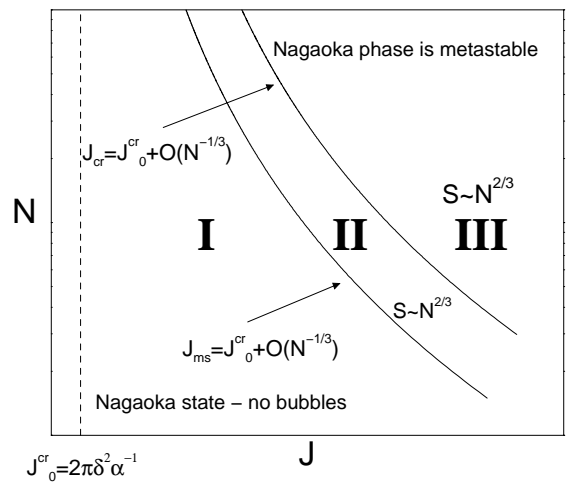}
\caption[]{Schematic phase diagram in the vicinity of the breakdown of
Nagaoka phase. Region I - Nagaoka phase is stable. Region II - Bubbles
bigger than the critical value are metastable. Region III - Bubbles
bigger than the critical value are stable.}
\label{f4}\end{figure}

The picture emerging from these calculation is as follows 
(see also Fig~\ref{f4}). At 
$J=\jcr$ (Eq. (\ref{jcr})) the Nagaoka state breaks down and one large 
AFM bubble with the area of order of $O(N^{2/3})$ is formed (Eq. (\ref{s0})). 
As the size of the bubble scales only sub-linearily
with system size, the magnetization per site is continuous
in the thermodynamic limit. In the vicinity of the transition 
point, only these large bubbles are stable, and therefore 
a large fluctuation is needed to destroy the Nagaoka phase. It is therefore 
a meta-stable phase. For larger $J$, $s_0$, the critical size for 
a stable bubble, decreases, and for $j\sim O(1)$ (i.e. $J-\jcr\sim\jcr$) 
it becomes finite and system-size independent. 
This behavior is typical for a first order phase
transition, where a domain of a new phase has to be large enough to survive.
However, 
the jump in the magnetization is not extensive. 

Why doesn't the system create many more bubbles, 
once the first bubble is formed, thus reducing the magnetization 
even further ? The reason is, as can be shown in Eq. (\ref{energy}), that the
kinetic energy is not a linear function of $s$. Therefore, the
energy cost to create two bubbles is more than twice the cost
to create one. On the other hand, the magnetic energy gain, is linear in $s$,
and therefore it would just double by the creation of the second 
bubble. 

There is an ongoing discussion in the literature regarding the existence
of a striped ground state in the Hubbard and $t-J$ models. Experimentally,
there is evidence for stripe modulations in doped cuprates \cite{expe},
which are generally believed to be described by the $t-J$ model.
Some authors found that the $t-J$ model ground state is indeed
striped for a wide range of doping \cite{ws}, while others claim
that uniform or phase separated states have lower energy \cite{hm}. 
According to the latter view, the origin of the experimental observation 
is attributed to
the competition between the local tendency for phase separation and the
long-range Coulomb interaction, which is missing in the
$t-J$ model \cite{c14}. 

In this study, we compared the energy of a striped state with
the the energy of a phase-separated state with an AFM bubble.
Within our approach, the kinetic energy increases in the striped
phase due to the higher surface energy, with nothing else to
compensate for this increase.  
Accordingly, the striped state energy is higher, suggesting that 
the long-range Coulomb repulsion is needed in order to create a striped 
ground state for $J/t\ll 1$. Our approach can not rule out the possibility
of a striped ground state of the $t-J$ model at $J\sim t$.

In conclusion, we presented analytical arguments, 
exact diagonalization results, and semi-classical calculations 
of the 2D $t-J$ model, which suggest that at small hole concentration,
$\delta$, and rather weak AFM coupling, $J$,
the Nagaoka ferromagnetic state becomes unstable towards the creation
of an AFM bubble. In this phase separated state, the
holes are confined to the FM regime. At the transition
only a single large $O(N^{2/3})$ bubble is (meta)stable.
Thus, the magnetization is continuous at the transition in the thermodynamic
limit. However, the jump in magnetization per unit area in a system 
with finite number of sites $N$, scales as $N^{-1/3}$.
This dependence of the critical bubble size, and thus of the magnetization,
on the size of the system is a typical finite size effect in
a phase separated ground state.

\bigskip 
\begin{acknowledgments}
We thank S.A. Brazovskii, A.V. Chubukov, Y.M. Galperin, and
S.A. Kivelson for helpful 
discussions and comments. The work at Princeton University 
was supported by the ARO and DARPA.
\end{acknowledgments}


\end{document}